\documentclass[conference,letterpaper]{IEEEtran}

\usepackage{amsmath}
\usepackage{comment}
\usepackage{stfloats}
\usepackage[dvips]{graphicx}
\usepackage{url}
\usepackage{times}
\usepackage{dsfont}
\usepackage{subfigure}
\usepackage{cite}
\usepackage{upgreek,textgreek}
\usepackage{color}
\graphicspath{{./figures/}}

\usepackage{tikz}
\usetikzlibrary{arrows, decorations.markings}

\renewcommand{\baselinestretch}{0.991}

\IEEEoverridecommandlockouts
\newcommand\copyrightnotice{%
\begin{tikzpicture}[remember picture,overlay]
\node[anchor=north,yshift=0pt] at (current page.north) {\fbox{\parbox{\dimexpr\textwidth-\fboxsep-\fboxrule\relax}{
\footnotesize \textcopyright 2021 IEEE. This paper has been accepted for presentation at UComms 2021. Personal use of this material is permitted.
Permission from IEEE must be obtained for all other uses, in any current or future media,
including reprinting/republishing this material for advertising or promotional purposes,
creating new collective works, for resale or redistribution to servers or lists,
or reuse of any copyrighted component of this work in other works.}}};
\end{tikzpicture}
}

\begin{document}

\title{An Event-Based Stack For Data Transmission Through Underwater Multimodal Networks \vspace{-2mm}}

\author{%
    \IEEEauthorblockN{\large Roberto Francescon, Filippo Campagnaro, Emanuele Coccolo,  \\[1mm]
    Alberto Signori, Federico Guerra, Federico Favaro, Michele Zorzi\vspace{1mm}}
    \vspace{-7.5mm}
    \thanks{%
        F.~Campagnaro (Corresponding author, email: campagn1@dei.unipd.it), E.~Coccolo, A.~Signori, F.~Guerra, F.~Favaro and M.~Zorzi are with the Department of Information Engineering, University of Padova, Italy. R.~Francescon is with Wireless and More srl, Padova, Italy.
        
        This work has been supported in part by the Italian Ministry of Education, University and Research (MIUR), and ERA-NET Cofound MarTERA (contract 728053). We would like to thank the smartPORT research group from TUHH for providing three acoustic modems and support.
    }%
}
\maketitle

\copyrightnotice

\begin{abstract}
The DESERT Underwater framework (http://desert-underwater.dei.unipd.it/), originally designed for simulating and testing underwater acoustic networks in sea trials, has recently been extended to support real payload data transmission through underwater multimodal networks. Specifically, the new version of the framework is now able to transmit data in real time through the EvoLogics S2C low-rate and high-rate acoustic modems, the SmartPORT low-cost acoustic underwater modem prototype (AHOI) for IoT applications, as well as Ethernet, surface WiFi, and the BlueComm optical modem. The system can also be tested in the lab by employing a simulated channel, and the EvoLogics S2C DMAC Emulator (DMACE).
\end{abstract}


\section{Introduction}\label{sec:intro}

Underwater networks allow a wide range of submarine applications, such as monitoring of oil spill, remote control of underwater vehicles, and coastline protection. 
In order to verify whether the performance of underwater networking protocols obtained through analysis and simulation can effectively be achieved in a real scenario, dedicated field evaluations must be carried out, in order to deal with the actual capabilities and limitations of real hardware.
Moreover, the best evaluation possible should be performed by transmitting real data instead of synthetic data automatically generated by the network simulator itself, in order to verify whether or not a certain application can actually be supported, at a certain level of performance, with the designed network. 
To achieve this, employing a framework that can easily allow to test a protocol stack both in simulations and in sea trials is essential. Furthermore, the same framework can become a tool to send real data through the underwater network. 
In recent years, several tools ~\cite{SUNSET_v2,aqua-net-mate2013,unetstack,cmre_cca} have been developed to support both simulations and sea trials. Among others, the SUNSET framework~\cite{SUNSET_v2}, based on the MIRACLE extensions~\cite{miracle-eurasip} of the network simulator ns2~\cite{ns2}, provides support for simulation, emulation and experimentation. 
Similarly, the Aqua-Net framework~\cite{aqua-net-mate2013} provides the possibility to switch between simulations and real-life experiments in underwater acoustic networks. The Aqua-Sim~\cite{aquasim} simulator, recently ported from ns2 to ns3, is able to exploit the emulation capabilities of ns3 to be employed for experimentation as well. Also the UAN framework~\cite{uan_ns3} is based on ns3, but it supports only network simulations.
The third version of the UnetStack framework~\cite{unetstack} provides not only the capabilities to run a software defined modem in a laptop, but also a set of APIs to develop real applications and therefore send real data through acoustics. The UnetStack software defined modem mode exploits the SubNero~\cite{subnero} acoustic modem features, while its simulation mode models the performance of the SubNero modem with high reliability.
The CMRE CCA Framework~\cite{cmre_cca} envisions a modular and multimodal communications architecture, designed to enable the deployment of advanced autonomous underwater solutions making use of smart, adaptive and secure underwater networking strategies. Mainly developed to be a reliable and robust framework for in-field activities, it also provides the possibility to perform network simulations, interacting with the CMRE channel simulator.
Another tool mainly developed for in-field activity is the EviNS framework~\cite{evins}, that can work either with S2C EvoLogics modems~\cite{S2CR} or with the EvoLogics DMACE real-time emulator~\cite{dmac_emul}. The latter employs a simulated underwater channel, specifically designed to map the EvoLogics modem performance. 
The open-source DESERT Underwater~\cite{DESERT_ucomms} framework, available online~\cite{DESERT}, offers a complete set of underwater networking capabilities, 
with the support of modular stacks and cross-layer communications.  Based on the MIRACLE extensions~\cite{miracle-eurasip} of the network simulator ns2~\cite{ns2}, it grants the co-existence of multiple modules within each layer of the protocol stack. 
The DESERT Underwater framework can support multiple and heterogeneous protocol stacks in each node. 
This feature enables the possibility to send packets through high frequency and low frequency acoustics as well as optical physical layers. The physical layers can be simulated~\cite{DESERT_ucomms}, emulated through maps of existing devices (e.g., via a Lookup Table), or connected with real modems to perform actual transmissions.
DESERT is the only open source framework able to evaluate multimodal optical and acoustic underwater networks through realistic simulations and sea trials.

In this work, we present the new design of the drivers used for in-field experiments with the DESERT Underwater framework developed at the University of Padova.
First, in Section~\ref{sec:drivers}, the new structure of the DESERT drivers is presented. This includes the general structure of the interface between the DESERT physical layer and the modems (Section ~\ref{sec:general_drivers}), as well as the specific implementation of the interface of two different acoustic modems, i.e., the S2C EvoLogics~\cite{S2CR} (Section ~\ref{sec:evo_drivers}) and the AHOI acoustic modems~\cite{renner_tosn} (Section ~\ref{sec:ahoi_drivers}), and of an end-to-end socket-based modem (Section ~\ref{sec:csa_drivers}), that can be used either to mock actual transmissions in the lab or to transmit data through Ethernet-based modems, such as the Sonardyne BlueComm optical modems~\cite{bluecomm_sonardyne}. 
In Section~\ref{sec:use_case}, we present different use cases and evaluate the drivers performance with an in-lab test campaign. 
In Section~\ref{sec:conclusions} we draw conclusions and illustrate ongoing work.

\begin{figure}[t]
      \centering
      \includegraphics[width=0.6\columnwidth]{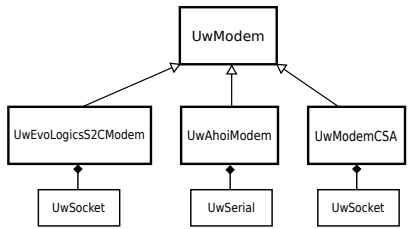}
      \caption{Structure of the inheritance relationships of the $\mathtt{UwModem}$ class.}%
      \label{fig:uwmodem}
\end{figure}

\section{Modem Drivers of DESERT Underwater}\label{sec:drivers}
The drivers provided by DESERT Underwater are responsible for managing real modems, allowing the transmission and reception of real data, as well as for configuring the devices according to the technical manual provided by the manufacturer.
Each modem driver is an ns2 module representing a physical layer in the ns2 architecture.


\begin{figure}[t]
      \centering
      \includegraphics[width=0.80\columnwidth]{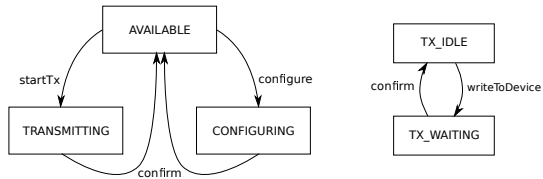}
      \caption{State machine for the DESERT drivers: general state machine $S$ (left-hand side) and transmission flow state machine $S_{TX}$ (right-hand side).}%
      \label{fig:modem_states}
\end{figure}


\subsection{Drivers General Structure}
\label{sec:general_drivers}
The general structure, presented in Fig.~\ref{fig:uwmodem}, includes the \texttt{UwModem} class, whose main task is to make available the basic functionalities common to each modem, such as transmission, reception and configuration. This is achieved through the use of a connector (\texttt{UwConnector}): in this context, connector refers to the 
physical interface connecting the communication hardware to the computer running DESERT Underwater, such as an Ethernet connection (either TCP or UDP) or a serial connection; to this aim, inheriting from \texttt{UwConnector}, the classes \texttt{UwSocket} and \texttt{UwSerial} implement the socket and the serial interfaces, respectively.

In addition to the main thread executing the event scheduler, each modem starts a transmission and a reception thread: these threads are created by the derived class. In the main thread the packets received from the upper layer are stored in a thread-safe transmission queue before being sent through the modem with the transmission thread. Similarly, packets received from the modem by the reception thread are stored in a thread-safe reception queue, periodically checked by the main thread which forwards the received packets to the upper layers. The checking period $TO_{RX}$ can be set in the simulation script. The lower $TO_{RX}$ the quicker the synchronization between the event scheduler and the real packet reception, but the higher the CPU use. For instance, in a laptop with an Intel core i7-8700
, $TO_{RX} = 10$~ms provides an average synchronization delay of 5~ms, and a CPU use (single core) of 4\%.

To correctly parse notifications from the modem and format commands, if any, a dedicated class, called \textit{interpreter}, is implemented in the derived class.

This structure allows for a simple integration of additional drivers for other devices: a new driver for a modem connected through the serial interface would be derived from \texttt{UwModem}, including the connector \texttt{UwSerial} and a custom state machine, used to send commands to the device, according to its specifications. Each event involving the modems, such as a packet to be transmitted or a notification from the device, is converted into an ns2 event and queued, along with all other simulator events, exploiting the ns2 event manager.

\subsection{S2C EvoLogics Drivers}\label{sec:evo_drivers}
EvoLogics GmbH~\cite{S2CR} manufactures commercial off-the-shelf devices for a wide range of environments and with various performance levels. These modems employ a technology called S2C (Sweep-Spread Carrier) which effectively tries to compensate multi-path. 
These devices can be accessed through an Ethernet interface, thanks to a TCP server listening for incoming connections.
The S2C devices are used, by the DESERT framework, in \textit{Command Mode}, which makes it possible to transmit and receive data and configure the modem easily, using only a TCP connection.

The EvoLogics devices have two types of packet transmissions: \textit{Instant Messages} (IM) and \textit{Burst} transmissions~\cite{KebkalSantander}. While IMs must be issued one at a time to not override the ongoing transmission and they exploit the lowest possible bitrate (976 bit/s for all devices with the exception of the S2C M HS, which provides a minimum bitrate of 1952 bit/s), burst transmissions exploit the full capability of the modem by determining the highest possible bitrate given the channel conditions estimated by the device itself. This mode works by storing all the issued transmissions in the modem's internal buffer, and then sending them as fast as possible. In our drivers, we implemented a state machine that tightly controls when packets are sent: a packet can be sent to the modem only if the confirmation of the previous transmission attempt has been received. This behavior implements in DESERT a stop and wait transmission policy, that is beneficial for the correct transmission of data and the control of the data link layer, but may rapidly deteriorate the performance, e.g., when large delays are involved. Indeed, the design choice of DESERT's drivers focuses more on the networking capabilities than on relying on specific abilities of the surrounding hardware to achieve the best performance on a single link.

The state machine for the S2C drivers includes two types of states: $S_{TX}$ represents the possible states for the transmission flow, specifically, \texttt{TX\_IDLE} and \texttt{TX\_WAITING}, and $S$
represents  the  general  state  of the  whole  device,  and can be \texttt{AVAILABLE}, \texttt{TRANSMITTING} or \texttt{CONFIGURING} to indicate whether the  modem  is  ready or busy  for  a  pending  transmission or configuration (Figure~\ref{fig:modem_states}). 
The main task of the state machine is to determine when a new configuration command or a new packet can be sent to the device, as the S2C modems can execute one command at a time, and the next command cannot be issued until the first is confirmed. The S2C modems make use of the {UwSocket} interface.

\subsection{AHOI Drivers}\label{sec:ahoi_drivers}
The AHOI modems are cheap acoustic underwater modem prototypes developed by the smartPORT group at TUHH (Hamburg)\cite{ahoioceans2019} and are accessed via serial connection. There is a practical interface, written in Python, that allows the use of the modems as a \textit{chat-like} application: this and other software, to manage the devices, can be found in the TUHH repositories\cite{ahoi-repo}. 
The AHOI modem is a cost-effective, high-frequency modem with a maximum bitrate of 260 bit/s (using Hamming encoding) and a maximum range of 150 m. It is intended mainly to be integrated in micro-AUVs and underwater sensor nodes. The modem operates, by default, in the 50-75 kHz band.

Also for the AHOI drivers, we keep track of the general modem state $S$, that can be \texttt{AVAILABLE}, \texttt{TRANSMITTING} or \texttt{CONFIGURING}: they specify if the modem is ready for new commands, is already transmitting or is being configured and they assure a correct transmission of all the packets, respectively. The transmission flow state $S_{TX}$, instead, can take the values \texttt{TX\_IDLE} and \texttt{TX\_WAITING} (Figure~\ref{fig:modem_states}). Also for the AHOI modems one command can be issued at a time, making use of the {UwSerial} interface. 

\subsection{Client-Server Architecture (CSA) Drivers}
\label{sec:csa_drivers}
These drivers are an end-to-end socket connection from a host to another remote host.
In contrast to S2C and AHOI drivers, CSA drivers make use only of the simple two-state (\texttt{TX\_IDLE} and \texttt{TX\_WAITING}) machine $S_{TX}$.
They make use of the {UwSocket} interface, and a simple two-state machine (AVAILABLE and BUSY). This driver implementation does not introduce further mechanisms for packet retransmissions, but only transmits a single packet as soon as the modem is in the \texttt{TX\_IDLE} state. 
These drivers can be used both to connect two hosts using either TCP or UDP through Ethernet or WiFi links, and with modems that make available an end-to-end Ethernet link, such as the Sonardyne BlueComm modems~\cite{bluecomm_sonardyne}.

\section{Use Case Examples and Performance Evaluation}
\label{sec:use_case}

In this section we investigate how the 
introduction of a state machine to control the use of the modems and 
ensure the synchronization between actual transmissions and the ns2 event scheduler affects the communication performance.


The application layer used to perform the following tests is \texttt{uwApplication}: this module enables the transmission through DESERT 
of every kind of stream through TCP or UPD sockets: from a chat-like application to the transmission of large files and images. This module is particularly useful as it allows the transmission of real data instead of mocked bytes. 

\begin{figure}[t]
      \centering
      \includegraphics[width=0.6\columnwidth]{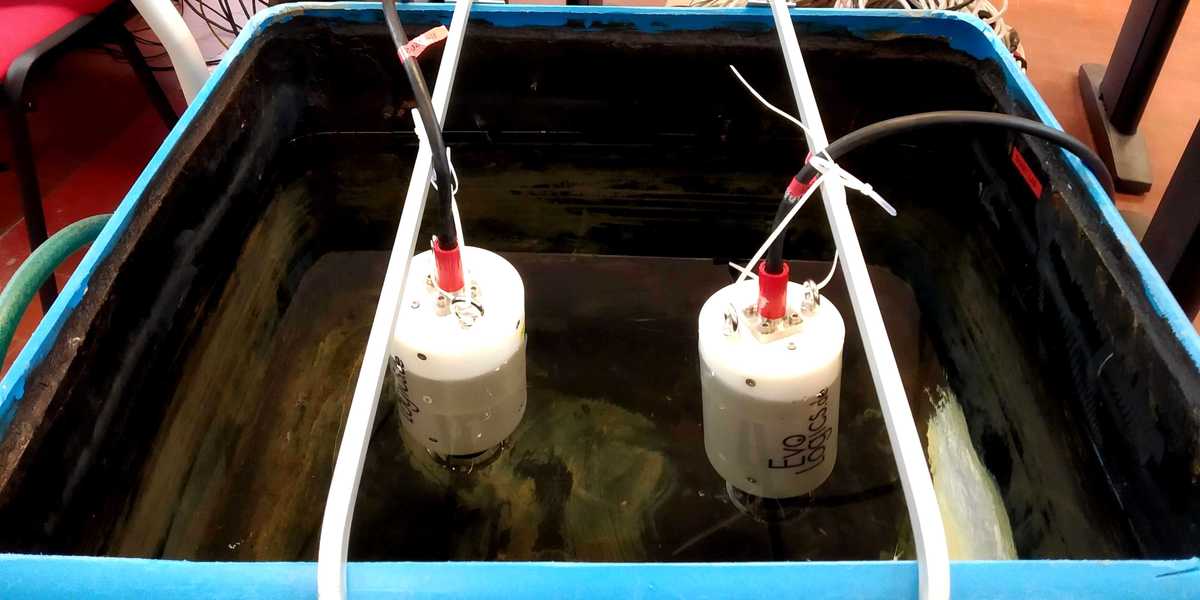}
      \caption{Setup of the test, for the S2C devices, in the insulated tank}
      \label{fig:tank-s2c}
\end{figure}

\begin{figure}[t]
      \centering
      \includegraphics[width=0.6\columnwidth]{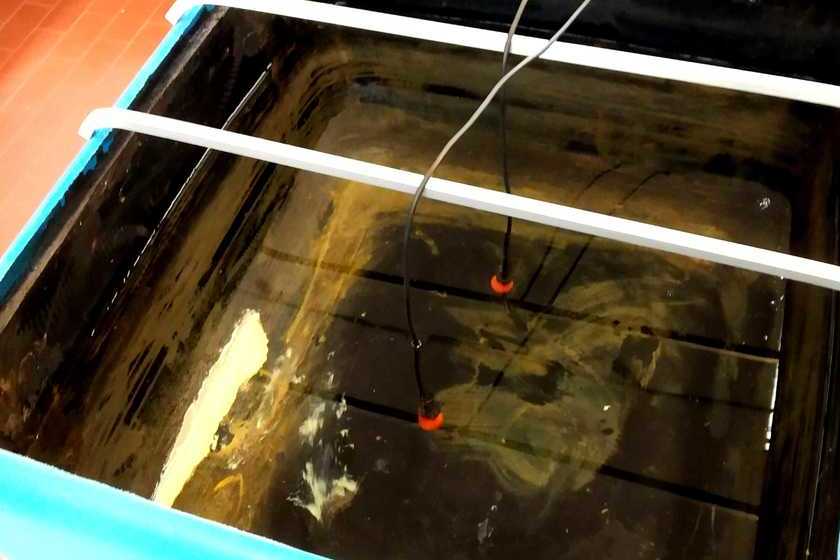}
      \caption{Setup of the test, for the AHOI modems, in the insulated tank}
      \label{fig:tank-ahoi}
\end{figure}

\begin{figure}[t]
      \centering
      \includegraphics[width=\columnwidth]{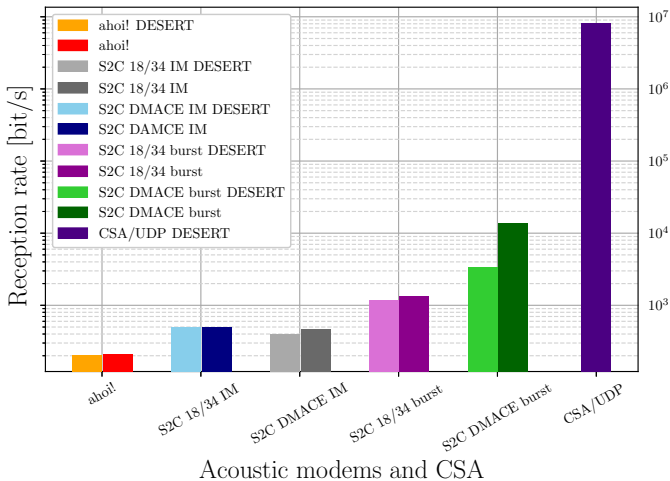}\vspace{-2mm}
      \caption{Average bitrate achieved while transmitting an image with various acoustic modems and the UDP end-to-end link.}
      \label{fig:multi-bitr}
\end{figure}

\subsection{End-to-End Image Transmission}
\label{sec:use_case_image}
To evaluate the performance of the DESERT drivers, we set up the transmission of a jpeg image of 2345 bytes from
one PC to another by employing different modems, 
using the devices both through DESERT and directly. In both cases we measured the datarate through the physical layer of the receiver stack $R_{RX}$ and the packet delivery delay $P_{dd}$, computed over 5 transmissions of the full image.

We tested the S2C modem drivers by employing both the DMACE emulator~\cite{dmac_emul} of the S2C 48/78 modems and two real S2C R 18/34 modems immersed in an insulated tank~\cite{SIGtank} (Fig.~\ref{fig:tank-s2c}). The same tank has been used to test the performance of the drivers developed for the AHOI modems (Fig~\ref{fig:tank-ahoi}), where the 2 transducers were deployed a few centimeters apart.

Fig.~\ref{fig:multi-bitr} shows that the performance loss caused by the use of the DESERT stack %
in the data transmission with AHOI and S2C modems in IM mode can be assumed to be negligible, especially if we consider the benefits of employing a protocol stack able to provide multimodal and multi-hop networking capabilities~\cite{DESERT_ucomms}.
Specifically, with the AHOI modems and the DESERT Underwater protocol stack, $R_{RX}$~=~205~bit/s and $P_{dd}$~=~1.40~s, 
while using the transmission tools provided by TUHH~\cite{ahoi-repo}, $R_{RX}$~=~212~bit/s and $P_{dd}$~=~1.315~s.
In both cases the packet size was set to 32 bytes. 
With the S2C 18/34 IMs through DESERT, instead, we were able to reach $R_{RX}$~=~494~bit/s and $P_{dd}$~=~1.125~s, while 
when transmitting without DESERT $R_{RX}$~=~500~bit/s and $P_{dd}$~=~1.04~s.
In this case the packet size was 64 bytes and the modems were deployed a few centimeters apart. 
By using the DMACE emulator, with the packet size still equal to 64 bytes, $R_{RX}$~=~400~bit/s with DESERT ($P_{dd}$~=~1.40~s) and 469 bit/s without DESERT ($P_{dd}$~=~1.15~s). This gap is due to the presence of a VPN to access the DMACE emulator, with a ping delay of 32 ms: the performance of the test without DESERT was not affected by this delay because we were just transmitting consecutive messages through the modem without checking its state. In this case the transmission period was manually tuned to maximize the reception rate. 



Using the S2C 18/34 devices for burst transmission, we measured $R_{RX}$~=~1170~bit/s and $P_{dd}$~=~2.69~s through DESERT and $R_{RX}$~=~1340 bit/s and $P_{dd}$~=~2.40~s when using the modem without DESERT. In this setup the modem bitrate was 3246 bit/s and the maximum packet size 512 bytes,
 while, transmitting burst data through the DMACE, with the nodes deployed 5 m apart, a packet size of 824 bytes was used
: in this last case we reached $R_{RX}$~=~3.385~kbit/s and $P_{dd}$~=~0.82~s through DESERT and $R_{RX}$~=~13.504~kbit/s and $P_{dd}$~=~0.75~s when using the modems directly. The bitrate of the emulated modem was 31.2 kbit/s.
In these two cases the performance difference is no longer negligible, because while the modems themselves buffer the outgoing packets and optimally transmit them, with DESERT we wait for the confirmation of each packet to be correctly received to have a software control of the MAC layer, as explained in Section~\ref{sec:evo_drivers}.


To prove that DESERT can also support the datarate transmission of a Bluecomm 200 optical modem (whose actual payload datarate varies from 0.25 Mbit/s to 8 Mbit/s), in Fig.~\ref{fig:multi-bitr} we also evaluated the maximum datarate achievable with the CSA drivers when transmitting UDP traffic (indeed, 8 Mbit/s).

\begin{figure}[t]
%
      \centering
      \includegraphics[width=0.55\columnwidth]{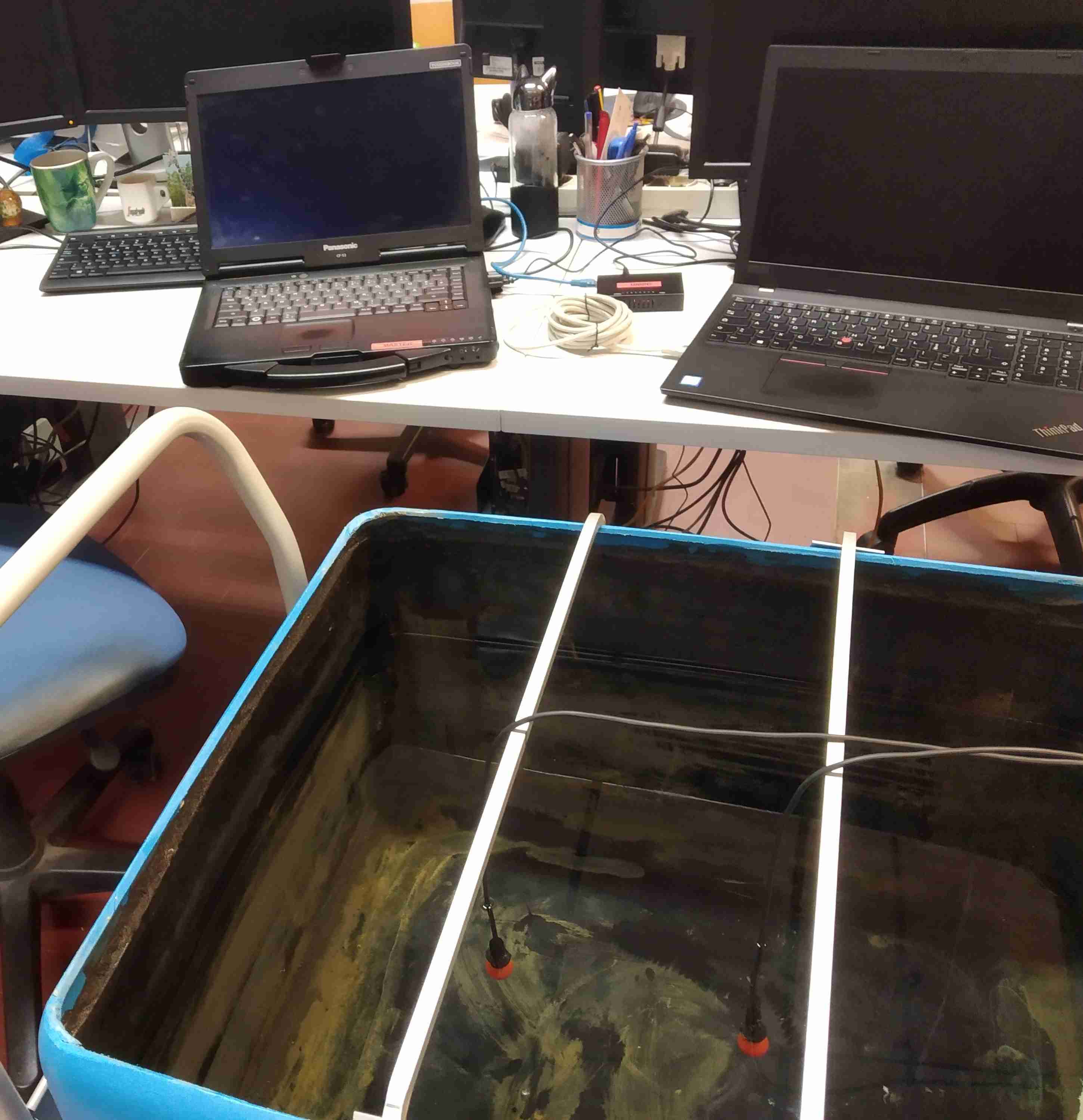}
      \caption{Setup of the test, for the multimodal network evaluation}
      \label{fig:multi_setup}
\end{figure}
\begin{figure}[t]
      \centering
      \includegraphics[width=0.60\columnwidth]{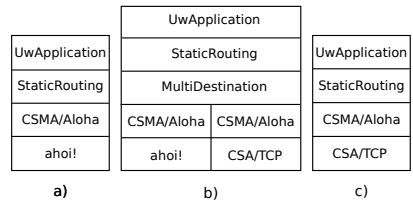}
      \caption{Stacks of the 3 types of nodes involved in the multimodal data muling test: a) UW nodes, b) buoy node and c) onshore node}
      \label{fig:nodes-stacks}
\end{figure}

\subsection{Multimodal two-hop underwater sensor network}
\label{sec:use_case_polling}
We implemented the two-hop network that is illustrated in Fig.~\ref{fig:nodes-stacks}
in order to test a multimodal network composed by three nodes, where node a) and node b) are connected through the AHOI modem, and node b) and node c) through a CSA/TCP link. 
Given that the CSA/TCP link is much faster than the AHOI acoustic link, we expect the bitrate to adapt to the latter, around 200 bit/s. The packet transmission from one stack to another, in node b), is performed by the \texttt{multi-destination} module~\cite{rizzoauvs} that operates between the MAC layer and the Network layer. 
This setup represents a situation in which an underwater data-collecting node wants to send data to a nearby buoy which is connected via a radio link to the onshore station.
Each of the three PCs illustrated in Fig.~\ref{fig:multi_setup} is running a separate DESERT instance: specifically, the workstation PC is running node a), and the two laptops are running node b) and node c), respectively.
The test performed confirmed that the end-to-end bitrate reaches the maximum exploitable bitrate of the AHOI modems, i.e., 205 bit/s, while $P_{dd}$~=~1.5~s.

\section{Conclusions and Future Work}
\label{sec:conclusions}
In this paper, we presented the new drivers structure of DESERT Underwater v3. The multimodal capabilities of DESERT allow now the real field deployment of a heterogeneous hybrid network, composed by EvoLogics acoustic modems, the low-cost AHOI acoustic modems, Bluecomm optical modems, and surface WiFi links. In this paper we provide an in-lab evaluation of the responsiveness of the framework compared to the direct use of the modems. We quantify the performance loss with each possible hardware configuration, proving that the system can be employed for real data transmission in multihop multimodal underwater networks, at the price of an almost negligible performance loss due to the DESERT processing time. Future work will focus on the evaluation of a complete multimodal network composed by both S2C and AHOI modems, as well as wireless surface links in real field tests to perform data gathering from underwater sensors~\cite{mdpi_polling} in the context of the RoBoVaaS project~\cite{robovaas} and the ability to optimally exploit modem-specific features such as the S2C burst data transmission.


\IEEEtriggercmd{\enlargethispage{-120mm}}
\IEEEtriggeratref{10}

\bibliographystyle{IEEEtran}
\bibliography{IEEEabrv,refen,biblio,biblio_Thesis_FC,misc}

\setlength{\textfloatsep}{12pt}

\renewcommand{\baselinestretch}{1}

\end{document}